 \documentclass[final,5p,times,twocolumn,authoryear]{elsarticle}

\usepackage{amssymb}
\usepackage{lipsum}
\usepackage{url}
\usepackage{amsmath}

\usepackage{xcolor}

\usepackage{lineno}

\journal{High Energy Astrophysics}

\begin{document}

\begin{frontmatter}



\title{Double White Dwarf Mergers as Progenitors of Long-Period Transients}
\author[1]{Manuel Malheiro}
\author[2]{Sarah V. Borges}
\ead{svborges@uwm.edu}
\author[3,4]{Jaziel G. Coelho}
\author[1]{Khashayar Kianfar}
\author[5,6]{Ronaldo V. Lobato}
\author[7]{Edson Otoniel}
\author[6,8,9,10]{Jorge A. Rueda}
\author[11]{Manoel F. Sousa}
\author[12,13]{Fridolin Weber}

\affiliation[1]{organization={Departamento de Física, Instituto Tecnológico de Aeronáutica},
addressline={Praça Marechal Eduardo Gomes 50},
city={São José dos Campos},
postcode={12228-900},
state={São Paulo},
country={Brazil}}

\affiliation[2]{organization={Department of Physics and Astronomy, University of Wisconsin-Milwaukee},
addressline={3135 North Maryland Avenue},
city={Milwaukee},
postcode={53211},
state={WI},
country={USA}}

\affiliation[3]{organization={Núcleo de Astrofísica e Cosmologia (Cosmo-Ufes) \& Departamento de Física, Universidade Federal do Espírito Santo},
state={Espírito Santo},
country={Brazil}}
\affiliation[4]{organization={Divisão de Astrofísica, Instituto Nacional de Pesquisas Espaciais},
country={Brazil}}

\affiliation[5]{organization={Centro Brasileiro de Pesquisas F\'isicas},
addressline={Rua Dr. Xavier Sigaud, 150},
city={Rio de Janeiro},
postcode={22290-180},
state={RJ},
country={Brazil}}

\affiliation[6]{organization={ICRANet},
addressline={Piazza della Repubblica 10},
city={Pescara},
postcode={65122},
country={Italy}}

\affiliation[7]{organization={Universidade Federal do Cariri, Instituto de Formação de Educadores},
country={Brazil}}

\affiliation[8]{organization={ICRA, Dip. di Fisica},
country={Italy}}
\affiliation[9]{organization={ICRANet-Ferrara},
city={Ferrara},
country={Italy}}
\affiliation[10]{organization={INAF},
country={Italy}}

\affiliation[11]{organization={Instituto de Física de São Carlos, Universidade de São Paulo},
addressline={Av. Trabalhador São-carlense 400},
city={São Carlos},
country={Brazil}}

\affiliation[12]{organization={Department of Physics, San Diego State University},
city={San Diego},
state={California 92182},
country={USA}}
\affiliation[13]{organization={Department of Physics, University of California at San Diego},
city={La Jolla},
state={California 92093},
country={USA}}

\begin{abstract}
 There is an ongoing discussion in the literature on the nature of long-period transients (LPTs), radio-emitting sources with periods ranging from hundreds to tens of thousands of seconds. Although some of these objects have been identified as white dwarf (WD) + M-dwarf binaries, this description currently does not fit the entire class. An example is GLEAM-X J162759.5--523504.3 (hereafter GLEAM-X J1627--5235), with a period of 1091~s, for which the lack of an optical counterpart disfavors the presence of such a binary system. In this case, GLEAM-X J1627--5235 could be interpreted as an isolated, massive, fast-rotating, and highly magnetized ($\sim 10^9$~G) WD pulsar. Its properties are consistent with a carbon–oxygen WD of mass $\sim 1.3\,\text{M}_\odot$ and radius $\sim 2500$~km, possibly supported by small-scale multipolar magnetosphere structures that keep it above the death line for WD-pulsars. We assess a double WD merger origin, modeling the post-merger rotational evolution under accretion, propeller, and magnetic braking torques. We find rotational age of $\sim 572$~Myr for GLEAM-X J1627--5235, i.e., the post-merger time required to reach its observed period. This result is consistent with current optical upper limits for GLEAM-X J1627--5235 and support the WD pulsar interpretation for this source.  We also discuss how the same model can apply to other LPTs.
\end{abstract}




\end{frontmatter}


\section{Introduction}\label{sec:1}

Since their discovery \citep{1968Natur.217..709H}, neutron star (NS) pulsars have attracted interest due to their extreme physics, including the densest matter known, intense magnetic fields, and rapid rotation. Their spin periods can reach milliseconds ($\Omega = 2\pi/P \sim \text{a few} \times 10^3,\mathrm{rad,s^{-1}}$), which implies both a deep gravitational potential $\sim (GM\Omega)^{1/3} \sim 0.1c^2$ and a compact radius of order $\sim15$ km. Pulsar emission, however, may not be exclusive to NSs: the WD in the binary AR Scorpii (AR~Sco), with a rotation period of $1.97$ minutes, and pulsed emissions across a wide spectrum of frequencies, has been proposed to be a WD-pulsar \citep{2016Natur.537..374M}. The spindown luminosity of AR~Sco exceeds by more than one order of magnitude the observed X-ray luminosity, as in traditionally rotation-powered pulsars, and the absence of accretion signatures suggests this classification as a plausible emission mechanism for AR~Sco~\citep{2017NatAs...1E..29B}. More recently, a counterpart to AR~Sco has emerged, identified within the system J191213.72--441045.1 (hereafter J1912--4410), exhibiting pulsed emission with a period of $5.30$~min~\citep{2023NatAs.tmp..120P}.

In this context, long-period radio transients (LPTs) are a new class whose properties place them in an ambiguous position relative to whether they are NSs or WDs~\citep[or even something else entirely, e.g.,][]{2024PhRvD.109f3004B, 2025ApJ...986...98Z}. They have emission periods ranging from 421 s \citep[CHIME J0630+25;][]{2024arXiv240707480D} to 6.5 hours \citep[ASKAP J1839-0756;][]{2025NatAs...9..393L}. Their nature has sparked considerable debate, as both NS and WD interpretations face significant challenges \citep[e.g.,][]{2022ApJ...940...72R,2022Ap&SS.367..108K,2022RNAAS...6...27L,2023Natur.619..487H}. For NSs, population-synthesis studies disfavor a population of isolated pulsars with spin periods of order $10^3$ s \citep{2024ApJ...961..214R}. For WDs, although coherent radio emission has long been predicted in theory, it had not been observationally established prior to the discovery of this class.

The discoveries of GLEAM-X J0704$-$37~\citep{2024ApJ...976L..21H} and ILT J1101+5521~\citep{2025NatAs...9..672D}, both confirmed WD+M-dwarf binaries, strengthened the case for them being WDs. In this case, they are potential WD pulsars and could have a similar emission mechanism to AR Sco and J1912--4410, in which the pulsed emission is likely amplified due to interactions of the WD magnetic field with winds coming from the companion~\citep{2016ApJ...831L..10G}, and not from curvature or synchrotron emission like for NS pulsars. Thus, binarity may be an essential feature in the observation of WD pulsars and may explain why LPTs with confirmed nature have been detected in WDs from binaries. Unfortunately, this binary WD pulsar model cannot explain the entire class. In particular, GLEAM-X J162759.5–523504.3 (hereafter, GLEAM-X J1627–5235) has optical MUSE/VLT observations that disfavor the presence of an emission exactly similar to LPTs known to be WDs ~\citep{2025MNRAS.538..925L}. More specifically, the optical limits place strong constraints on the presence of a WD+M-dwarf binary at the given distance, which could be reconciled if the distance to the source has been underestimated, but increasing the distance would further increase the radio luminosity, which is already significantly brighter than for WD+M-dwarf LPTs.

GLEAM-X J1627–5235 has a rotation period of 1091 s and a spindown rate upper limit $\dot{P} \leq 1.2 \times 10^{-9}$ s s$^{-1}$ \citep{2022Natur.601..526H}. Moreover, unlike confirmed binary LPTs, in which the few-hour period is likely orbital, the period of GLEAM-X J1627–5235 is too short for a WD+M-dwarf binary and is possibly the object spin period. Observationally, its radio emission shares characteristics with that of GLEAM-X J0704$-$37, including high linear polarization, short-duty-cycle bursts, and millisecond-scale microstructure. A NS interpretation for these long-period objects, however, would require nonphysical magnetic fields~\citep[$\gtrsim 10^{17}$ G,][]{2024ApJ...976L..21H}. If interpreted as an isolated WD, it can be a fast-rotating, highly magnetized remnant with a mass of $\sim 1.3$~M$_\odot$, period $\sim 10^3$ s, and magnetic fields $\sim 10^9$ G, consistent with double WD merger products. More importantly, due to its extreme parameters, this post-merger massive isolated WD can be above the death line for a WD-pulsar and does not need to rely on the interactions with the companion to have a pulsar emission.

Although the WD interpretation remains indirect due to the absence of optical counterparts or measured spindown rates, our modeling yields falsifiable predictions. If confirmed, the merger origin would support the hypothesis that strong magnetization is a natural outcome of the merger dynamics \citep{2012ApJ...749...25G}. These findings reinforce the scenario explored here, in which some LPTs may originate from double WD mergers, producing isolated, fast-rotating, highly magnetized WDs capable of coherent radio emission over gigayear timescales. 

The paper is structured as follows. In Section~\ref{sec:2}, we discuss the current population of observed WDs with similar parameters and the ongoing debate regarding their possible origin in double WD mergers. Because LPTs are radio emitters, much of the recent literature has focused on the so-called pulsar death lines and limits to the spindown luminosity~\citep[e.g.,][]{2022ApJ...940...72R,2023ApJ...943....3T,2023arXiv230714829T}. We revisit these radio emission constraints in Section~\ref{sec:3} to assess the physical conditions required for coherent radio activity in WD pulsars. In this discussion of the death-line, we include all LPTs in which this model can be applied. In Section~\ref{sec:4}, we examine the post-merger cooling evolution of GLEAM-X J1627–5235 and compare with the current upper limits from VLT/MUSE. Section~\ref{sec:5} presents our rotational evolution models based on a double WD merger scenario, incorporating accretion, propeller, and magnetic braking torques. As an example, we applied this model to two LPTs, GLEAM-X~J1627–5235 and GPM~J1839-10. In Section~\ref{sec:6}, we provide the maximum expected optical magnitudes for LPTs as isolated WDs, to guide future observations. Finally, in Section~\ref{sec:7}, we discuss and summarize our results.

\section{Observation of massive, fast-rotating, highly magnetic WDs and their post-merger origins}\label{sec:2}
\begin{table*}[h!]
\centering
\caption{Parameters of LPTs presented in Fig.~\ref{fig:deathlineWD} and some isolated WDs. Columns indicate rotation period ($P$), spindown rate ($\dot{P}$), magnetic field ($B$), mass ($M$), radius ($R$), temperature ($T$), age ($\tau$), distance ($d$), and extinction in the G-band ($A_G$). We have not included the distance and extinction for isolated WDs, because they are not relevant to this work. }
\vspace{2mm}
\begin{tabular}{cccccccccc}
\hline
Sources & $P$ (s) & $\dot{P}$ ($ 10^{-11}$~s.s$^{-1}$) & $B$ ($ 10^{8}$~G) & $M$ ($M_\odot$) & $R$ (km) & $\tau$ (Myr) & $d$ (kpc) & $A_G$ &Ref. \\
\hline \hline
\multicolumn{10}{c}{LPTs} \\
\hline
GLEAM-X J1627 & 1091.17 & $< 120$ & $\leq 6031^{a}$ & 1.33$^{c}$ & 2500$^{c}$ & $\geq 0.014^{d}$ & $1.3\pm0.5$ &$\sim$0.7  & (1)\\
GPM J1839-10 & 1318.20 & $< 0.036$ & $\leq 115^{a}$ & 1.33$^{c}$ & 2500$^{c}$ & $\geq 58.0^{d}$ & $5.7\pm2.9$& $\sim$3.4 & (2) \\
ASKAP J1935+2148 & 2225.31 & <12 & $\leq 2723^{a}$ & 1.33$^{c}$ & 2500$^{c}$ & $\geq 0.29^{d}$ & $\sim$4.85 &$\sim$1.8  & (3)\\
CHIME J0630+25 & 421.35 & 0.08 &$97^{a}$ & 1.33$^{c}$ & 2500$^{c}$ & $8.3^{d}$ & $0.17^{+0.31}_{-0.10}$ &$\sim$0.3  & (4)\\
DART J1832-0911 & 2656.24 & <0.9 &$\leq 815^{a}$ & 1.33$^{c}$ & 2500$^{c}$ & $\geq 4.7^{d}$ & $4.5\pm1.2$ &$\sim$5.8 & (5)\\
ASKAP J1755-2527 & 4186.32 & $<1$ & $\leq 1078^{a}$ & 1.33$^{c}$ & 2500$^{c}$ & $\geq6.6^{d}$ & $\sim$4.7 &$\sim$8.9 & (6)\\
\hline \hline
\multicolumn{10}{c}{Isolated WDs} \\
\hline
J2211+1136 & 70.32 & $-$ & 0.15$^{b}$ & 1.27 & 3210 & 2600-2900$^{e}$ & & & (7) \\
J1252-0234 & 317.28 & $-$ & 0.05$^{b}$ & 0.65 & 8374 & 1136 $\pm$ 95$^{e}$ & & & (8)\\
J1901+1458 & 416.24 & $<1$ & 6-9$^{b}$ & 1.33-1.36 & 2140 & 10-100$^{e}$ & & & (9)\\
J0317-0853 & 725.74 & $-$ & 1.7-6.6$^{b}$ & 1.34 & 2456 & 281$^{+36}_{-31}$~$^{e}$ & & & (10)\\
J0006+3104 & 1385 & $-$ & $\sim$2.5$^{b}$ & 1.06 & 4528 & 200$\pm 10^{e}$ & & & (11) \\
\hline
\end{tabular}

\vspace{2mm}
$^{a}$ Magnetic field from dipole braking: $B \sin{\theta} = \sqrt{\frac{3 c^3 I P \dot{P}}{8 \pi^2 R^6}}$, with $M=1.33,M_\odot$, $R=2500$ km, $I = \frac{2}{5} M R^2= 6.61\times 10^{49}$ g cm$^2$, and $\theta$ is the angle between the axis of rotation and the magnetic axis.
$^{b}$ Magnetic field inferred from atmosphere/spectral modeling. $^{c}$: fixed parameter from our model.
$^{d}$ characteristic age ($\tau_{\rm char} = P/2\dot{P}$),$^{e}$: cooling age from surface temperature. \
(1) \cite{2022Natur.601..526H},
(2) \cite{hurley-walker/2023},
(3) \cite{2024NatAs...8.1159C},
(4) \cite{2025ApJ...990L..49D},
(5) \cite{2025Natur.642..583W},
(6) \cite{2024MNRAS.535..909D},
(7) \cite{2021ApJ...923L...6K,2021MNRAS.503.5397K},
(8) \cite{2020ApJ...894...19R,2020MNRAS.499.2564G},
(9) \cite{2021Natur.595...39C},
(10) \cite{1997MNRAS.292..205F,2003ApJ...593.1040V,2010A&A...524A..36K},
(11) \cite{2026ApJ...998..292G}.
\label{tab:LPT_WDs}
\end{table*}

WDs can exhibit magnetic fields with intensities ranging from $10^3$~G to $10^9$~G \citep[see, e.g.,][for a review]{2009A&A...506.1341K,2010yCat..35061341K,kepler/2013,2015MNRAS.446.4078K,2023ApJ...944...56A,2015SSRv..191..111F}. They can also show rotational periods spanning from seconds to hours. Among the fastest rotation periods ever observed is that of LAMOST~J0240.51+1952, with an extraordinary rotation period of $24.93$~s~\citep{Pelisoli2022}, and CTCV~J2056-3014, with a period of $29.6$~s~\citep{LopesDeOliveira2020,2021A&A...656A..77O}. The presence of a strong magnetic field and rapid rotation provides strong evidence for past and current binary interactions. In such systems, the progenitors likely passed through a common-envelope evolution~(CEE) phase~, which is responsible for forming close binaries~\citep[see, e.g.][]{2019MNRAS.490.2550I, 2020cee..book.....I,2026ApJ...998...34B}. Subsequent episodes of mass transfer can both spin up the WD by transferring angular momentum and amplify magnetic fields via crystallization and rotation-driven dynamo processes~\citep{2021NatAs...5..648S}.
 
There is also a population of magnetic, isolated, and massive WDs that has been identified across several investigations \citep{2005A&A...441..689A,2007A&A...465..249A,2013ApJ...771L...2H,2016MNRAS.455.3413K,2017MNRAS.468..239C,10.1093/mnras/sty3016,2018MNRAS.480.4505J,2019A&A...625A..87C}. Notable examples include ZTF~J1901+1458, with $M \approx 1.31\,M_\odot$ and $B \approx (6.0\text{--}9.0)\times10^8$~G \citep{2021Natur.595...39C}, and PG~1031+234, which holds the record for the highest magnetic field detected, $B \approx 10^9$~G \citep{1986ApJ...309..218S,2009A&A...506.1341K}. In this case, double WD mergers, a more extreme form of binary interaction, have long been proposed as a natural explanation for the origin of these objects \citep{2000PASP..112..873W}. Previous studies have shown that magnetic WDs are, on average, more massive than their non-magnetic counterparts, which agrees with this merger origin interpretation~\citep{kepler/2013}

Theoretical work led by numerical simulations shows that mergers can form sub-Chandrasekhar WD central remnants \citep{1990ApJ...348..647B,2004A&A...413..257G,2009A&A...500.1193L, 2012A&A...542A.117L,2012ApJ...746...62R,2013ApJ...767..164Z, 2014MNRAS.438...14D,2018ApJ...857..134B}. A dynamo process and the high spin due to the total orbital momentum of the merger are expected to be responsible for magnetizing the WD \citep{2012ApJ...749...25G}. These mergers are expected to be numerous, with a local universe rate of $(3.7$--$6.7)\times 10^5$~Gpc$^{-3}$~yr$^{-1}$ \citep{2017MNRAS.467.1414M, 2018MNRAS.476.2584M,2022ApJ...941...28S}, which is $5$--$8$ times larger than that of SN~Ia \citep{2009ApJ...699.2026R}. Therefore, we expect many mergers do not produce SNe~Ia \citep{2020ApJ...891..160C}, some of them lead to metastable super-Chandrasekhar WDs that can form neutron stars by delayed gravitational collapse \citep{2018ApJ...857..134B}, and finally, some lead to massive, highly magnetized, fast-rotating WDs. 

The observed transverse velocities of massive WDs also hint that a
non-negligible fraction might have originated from double WD mergers
\citep{2020ApJ...891..160C}. A recent study \citep{kilic2023merger}
looked for possible merger remnants among the most massive $25$ WDs in a sample of $100$~pc from the \textit{Montreal WD Database}, analyzing features such as rapid rotation, high tangential velocities, magnetic properties, and unusual atmospheric compositions, concluding that $14$ sources strongly suggest WD mergers or similar phenomena. \citet{toonen2017} elucidated an additional aspect that bolsters the merger hypothesis. Using WDs within a $20$ pc sample and recent data from the {\it Gaia} space observatory, they showed that the prevalence of WD binaries ($\sim$ 25$\%$) is lower when compared to the occurrence of double systems containing solar-type main-sequence stars ($\sim$ 50$\%$). This discrepancy is consistent with the idea that 10$\%$ to 30$\%$ of all isolated WDs result from coalescence processes \citep{toonen2017}.

In terms of rotation, isolated magnetic WDs can occupy two extremes, with spins that are either significantly faster or slower than those of their non-magnetic counterparts. At one end, a number of magnetic WDs exhibit extremely long periods ($>100$~years), far exceeding the average period of non-magnetic WDs~\citep[around a day,][]{2015SSRv..191..111F}. On the other hand, recent observations have revealed isolated WDs with very short rotation periods, ranging from seconds to minutes. In particular, SDSS~J221+1136 has a spin period of $P = 70.32$~s \citep{2021ApJ...923L...6K}, and ZTF~J1901+1458 has $P = 416.20$~s \citep{2021Natur.595...39C}. This diversity can be explained by the formation and evolution of post-merger WDs. These objects are initially born as rapid rotators; however, magnetic braking and interactions with the massive disk formed shortly after the merger can efficiently remove angular momentum, in some cases producing periods longer than those of typical isolated non-magnetic WDs. Ultimately, the rotation period of an isolated magnetic WD can span a wide range, depending on its initial period, age, magnetic field strength, and disk properties. Consistent with this picture, recent simulations of the rotational and thermal post-merger evolution of WDs have shown that observed isolated systems, in particular SDSS~J2211+1136 and ZTF~J1901+1458, can be reproduced as products of WD mergers \citep{2022ApJ...941...28S}.

Table~\ref{tab:LPT_WDs} showcases a subset of isolated, fast, and magnetic WDs known in the literature. The rotation period, magnetic field, temperature, and mass have been derived from photometric and spectral analyses, as well as model-atmosphere features. Cooling ages are estimated using evolutionary and cooling models~\citep[see][, and references therein]{2021MNRAS.503.5397K,2021Natur.595...39C,2020MNRAS.499.2564G, 2010A&A...524A..36K}. In the case of ZTF J1901+1458, the WD radius is derived from photometric and distance measurements, and the mass is obtained from the mass-radius relation. For TMTS J0006+3104, the radius is determined from spectral energy distribution fitting. For the other WDs, the radius is estimated from the surface gravity, $g$, i.e., $R = \sqrt{G M/g}$. The spindown rate for these sources remains unknown, except for ZTF J1901+1458, which has an upper limit value for this parameter~\citep[see][]{2021Natur.595...39C}. 

Table~\ref{tab:LPT_WDs} also presents the observed spin, (upper limit) spindown, distance, and extinction in the G-band for GLEAM-X J1627--5235 and other LPTs that are potential isolated WDs. For the extinction, we considered the objects dispersion measures and the method to obtain $A_G$~(DM $\rightarrow N_H \rightarrow A_V\rightarrow A_G$) described in Sec.~\ref{sec:4}. We include all LPTs that have measured spindown limits, no confirmed companion, and periods in the order of $\sim 10^3$~s, which was chosen to guarantee that the minimum magnetic field for the object to be above the death-line of WD pulsars is around $10^9$~G. We also included the (upper limit) magnetic field, determined from the dipole braking formulae; the (lower limit of) the characteristic age; and the mass and radius, which are fixed parameters in our model.  We have not included the LPTs ILT~J1101+5521, GLEAM-X~J0704-36  and ASKAP~J1448-6856 due to their WD+M-dwarf binary nature~\citep{2025A&A...695L...8R,2025NatAs...9..672D,2025MNRAS.542.1208A}, GCRT~J1745-3009 due to the lack of spin down upper limits~\citep{2005Natur.434...50H}, ILT~J1634+44  because this object is spinning-up~\citep{2025ApJ...988L..29D}, and ASKAP J1839-0756 due to its high period~\citep{2025NatAs...9..393L}.

When we began this work, we included GPM~J1839$-$10 as a candidate isolated WD. However, it has recently been proposed to reside in a binary system, despite the lack of a confirmed optical counterpart, based on the detection of a secondary period of 8.75~h in radio in addition to the primary period of 1318~s~\citep{2025arXiv250715352H}. In this interpretation, the longer period corresponds to the orbital period, while the shorter period is a beat period between the orbital period and the WD spin period (1265~s). We opt to keep this object in our sample with appropriate caveats. We first note that GPM~J1839$-$10 may be a fast, massive, and magnetic WD in a binary system.REJ~0317$-$853, for example, is in a wide binary system, yet it is a strong candidate to be formed through a WD+WD merger due to its extreme parameters~\citep[see Table~\ref{tab:LPT_WDs}. Also][]{2010A&A...524A..36K}.

In close binaries, a post-merger WD can be present in several ways. One possibility is a CEE occurring in the still-triple system, in which two WDs are in a close binary orbiting a giant. In this case, the two WDs can merge during the event~\citep[see, e.g.,][]{2021MNRAS.500.1921G}. The final product would be a close binary with a massive WD and the core of the giant~(either another WD or a hot subdwarf). In general, the biggest indication of a merger is the potential high mass for the WD ($\gtrsim$1.3~$M_{\odot}$), given that high magnetic fields and fast periods can also be explained by other, less extreme binary interactions. Thus, until the origin of the secondary period is confirmed to be orbital and the nature and masses of the objects are determined, the merger nature of GPM~J1839$-$10 remains possible, so we retain this object in our analysis.

\section{Death lines for WD-pulsar}\label{sec:3}

We now recall the traditional death lines for pulsar radio emission and apply them to WDs. The radio emission models consider the acceleration of electrons and positrons ($e^\pm$) to extreme energies, such that the radiation originated by these particles could create new $e^{\pm}$, which in a sequence creates new ones, leading to a cascade. This avalanche is expected to produce coherent radio-wave flux. We refer the reader to \cite{2022ARA&A..60..495P} for a recent review. Therefore, traditionally, the minimum condition for pulsar radio emission is that the pulsar magnetosphere produces the first generation of pairs.

The basics of such a process roots in the \textit{polar-cap} model by \cite{goldreich/1969,sturrock/1971,ruderman/1975}. The electric field component along the \textit{open} magnetic field lines does not vanish in the pulsar \textit{gap}, accelerating charged particles to sufficiently high Lorentz factor to produce $\gamma$-rays by synchro-curvature radiation. Photons interact with the magnetic field, leading to the first generation of pairs, i.e., $\gamma+B \to e^+e^-$ \citep{sturrock/1971,ruderman/1975}. The electric potential drop along an open field line above the surface is $\Delta V = B_s \Omega\, h^2/c$ where $h\ll R$, being $h$ the gap thickness and $R$ the stellar radius~\citep[see Eqs. A7 in][]{ruderman/1975}). The condition for pair production by photon-magnetic field interaction is \citep{sturrock/1971,chen/1993}
\begin{equation}\label{line}
\frac{3}{2}\gamma^3\frac{\hbar}{2mcr_{c}}\frac{h}{r_{c}}\frac{B_{s}}{B_{g}}\approx\frac{1}{15},
\end{equation}
where $r_c$ is the curvature radius of the magnetic field lines, $B_g =m^2c^3/(e\hbar) = 4.42\times10^{13}$~$\rm{G}$, $B_s$ the magnetic
field on the surface, $\gamma = e \Delta V/(m c^2)$ the electron's Lorentz factor, and $m$ its rest-mass. With all the above, one can express the gap thickness and the electric potential drop in the gap as a function of the other pulsar parameters, i.e.,

\begin{align}
    h &= 4.76\times 10^6 \left( \frac{r_c}{10^6\,\text{cm}} \right)^{2/7}
        \left( \frac{P}{10^3\,\text{s}} \right)^{3/7} \notag \\
      &\quad \times \left( \frac{B_s}{10^9\,\text{G}} \right)^{-4/7}
        \,\text{cm}, \label{eq:h} \\[4pt]
    e \Delta V &=  1.42\times 10^{12} \left( \frac{r_c}{10^6\,\text{cm}} \right)^{4/7}
        \left( \frac{P}{10^3\,\text{s}} \right)^{-1/7} \notag \\
      &\quad \times \left( \frac{B_s}{10^9\,\text{G}} \right)^{-1/7}
        \,\text{eV}. \label{eq:eV}
\end{align}
The potential drop reaches its maximum value  $\Delta V_{\rm max} = B_s R^3 \Omega^2/c^2$ for the gap thickness of a magnetic dipole, which is given by the polar cap size, $r_p \approx R\,\theta_p$, being $\theta_p = \arcsin(\sqrt{R/r_{\rm lc}})\approx \sqrt{R/r_{\rm lc}}$ the polar cap angular size, i.e., the angular size from the rotation axis to the last open field line, with $r_{\rm lc} = c/\Omega$ the light cylinder radius. Therefore, $h_{\rm max} = r_p = R\sqrt{R/r_{\rm lc}}$. The condition that $h < h_{\rm max}$, or equivalently $\Delta V < \Delta V_{\rm max}$, sets the WD-pulsar's death line, so a lower limit to the very curved magnetic field for pair creation by photon-magnetic field interaction, i.e.,
\begin{equation}\label{eq:Bsdl}
\begin{aligned}
    B_s \geq &\,6.02\times 10^{10} \left( \frac{10^8\,\text{cm}}{R} \right)^{21/8}
    \left(\frac{r_c}{10^6\,\text{cm}} \right)^{1/2} \\
    &\times \left(\frac{P}{10^3\,\text{s}} \right)^{13/8} \,\text{G}.
\end{aligned}
\end{equation}

\begin{figure*}
    \centering
    \includegraphics[width=0.46\hsize,clip]{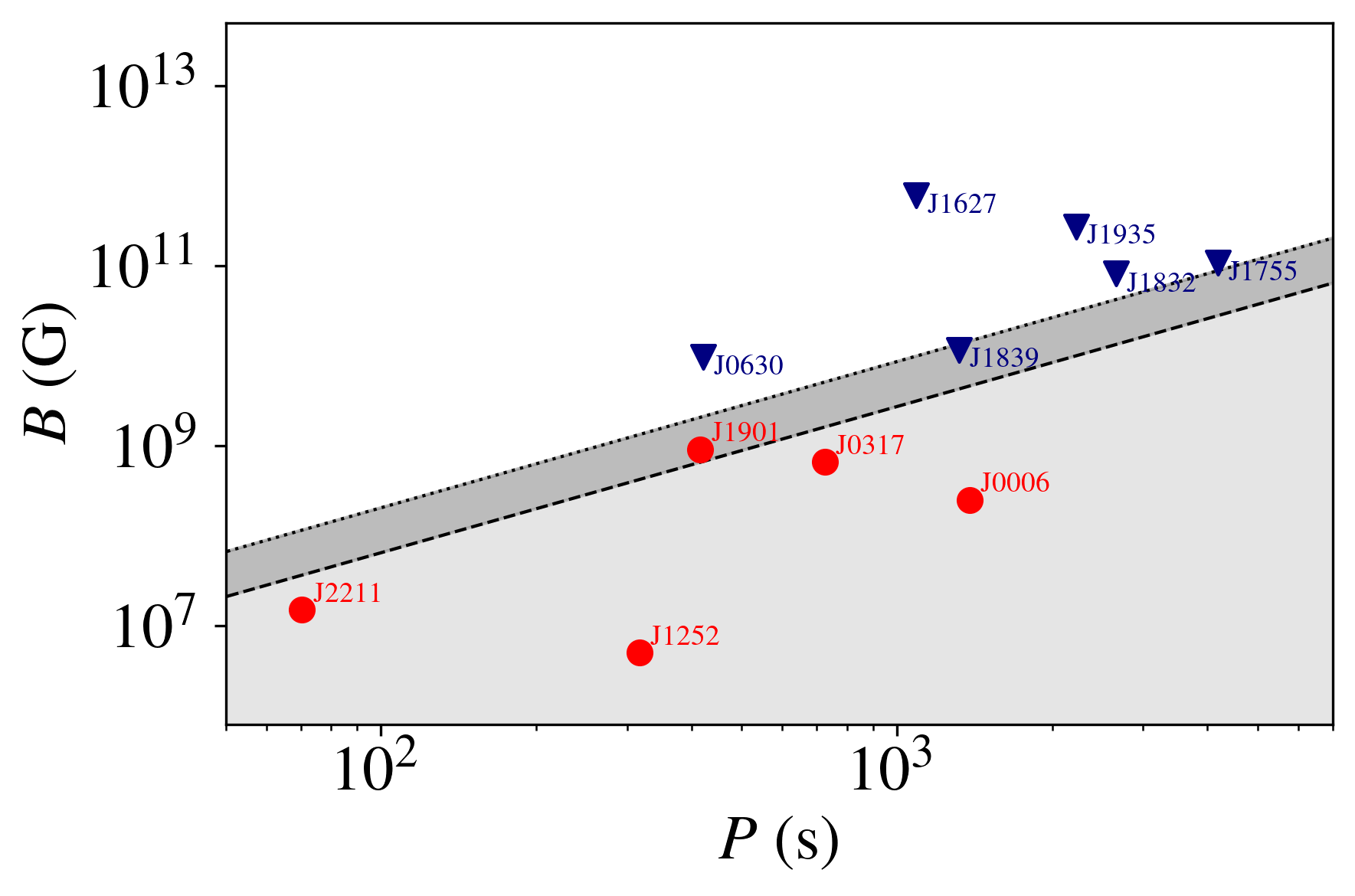}    
    \includegraphics[width=0.45\hsize,clip]{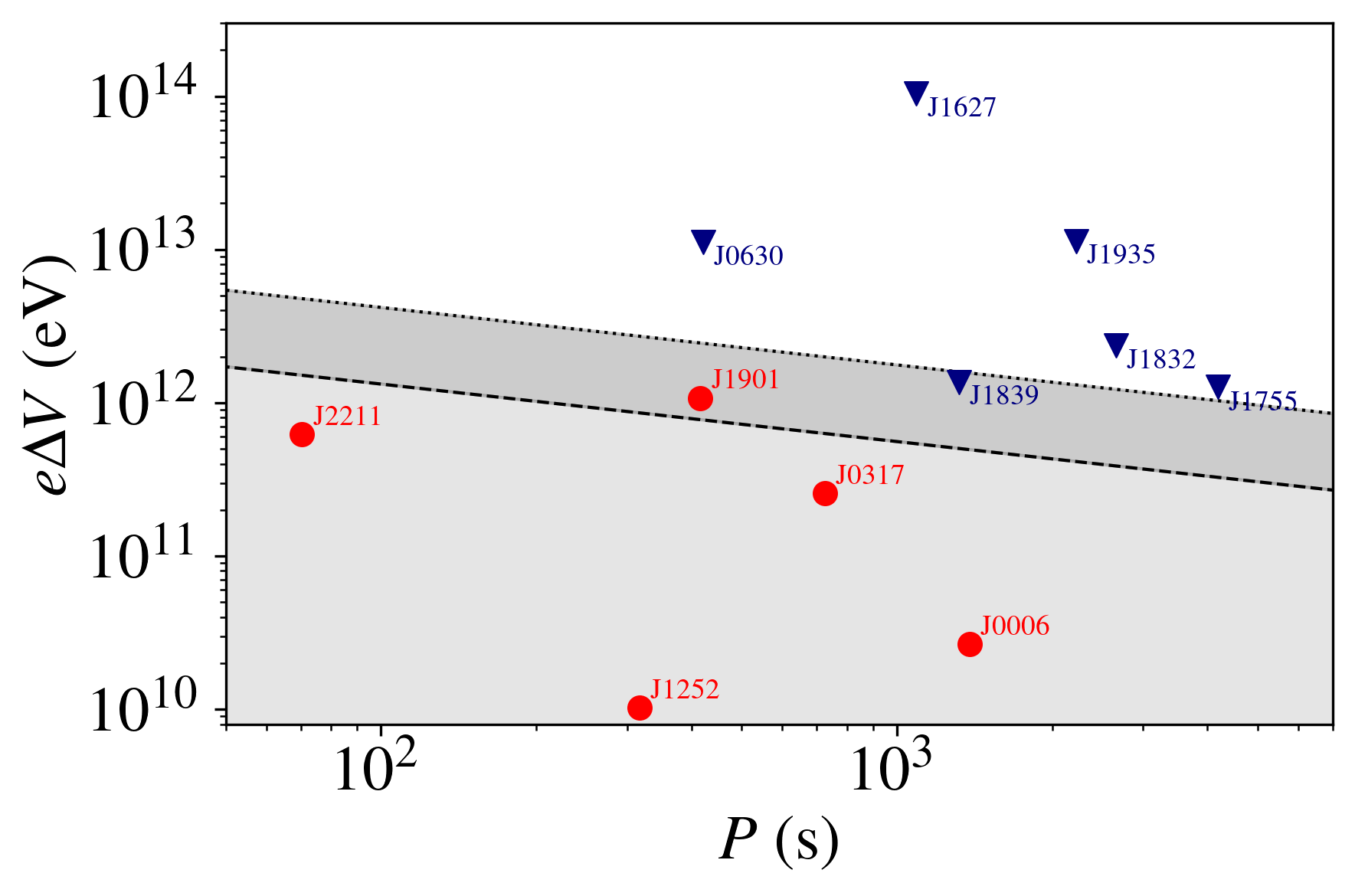}
    \caption{\textbf{Left:} Magnetic field limits set by the death line for WD-pulsars. Black-dashed curve: $r_c/R = 0.001$. Black-dotted curve: $r_c/R=0.01$. Other parameters are $R=2500$ km ($M\approx 1.3\,\text{M}_\odot$) and $h = h_{\rm max}=r_p$. Period and magnetic field values of magnetic WDs (red circles) are from Table~\ref{tab:LPT_WDs}, while the upper limits of magnetic field for LPTs (blue triangles) are from Table~\ref{tab:LPT_WDs}. Exceptionally, CHIME J0630+25 has a measured spindown rate, and thus the magnetic field is not an upper limit but an estimate based on dipole braking. \textbf{Right:} Electric potential drop limits set by the death line for WD-pulsars. Black-dashed curve: $r_c/R = 0.001$. Black-dotted curve: $r_c/R=0.01$. Other parameters are as in the left panel.}
    \label{fig:deathlineWD}
\end{figure*}

Figure \ref{fig:deathlineWD}.a shows the WD pulsar death lines set by Eq. (\ref{eq:Bsdl}) for $R=2500$ km ($M\approx 1.3\text{M}_\odot$), $h = h_{\rm max}=r_p$, and two values of the curvature radius, $r_c = 0.001 R$ and $r_c = 0.01 R$. Here $R$ is the WD stellar radius. Radii of curvature $r_c \ll R$ do not describe a global centered-dipole/open-line geometry, but rather local, small-scale, near-surface multipolar spot-like magnetic structures anchored in the outer layers of the WD. In such non-dipolar polar-cap regions, neighboring opposite-polarity flux bundles and strong higher-order multipoles components (or an offset dipole) can bend open lines on scales much smaller than $R$, even $r_c \sim 10^{-3}$--$10^{-2} R$, also enhancing pair creation (see, e.g., \citealp{gil/2001, 2011ApJ...726L..10H}, for discussions of this effect in pulsars). The source points are for the LPTs parameters (rotation period and magnetic field) in Table \ref{tab:LPT_WDs}. Figure \ref{fig:deathlineWD}.b shows the physical situation from the potential drop viewpoint, i.e., it plots the maximum potential drop $e \Delta V_{\rm max}$ using the death line shown in Fig. \ref{fig:deathlineWD}.a, and using the observed magnetic field value reported in Table \ref{tab:LPT_WDs}. The condition for radio emission imposes that the corresponding source lies above the gray-filled region.

Figure \ref{fig:deathlineWD} suggests constraints on the WD magnetosphere parameters for the radio emission of LPTs. The figure indicates the upper limits to the magnetic field of these sources given by the magnetic dipole braking model in Table \ref {tab:LPT_WDs}. It arises that, as expected, the magnetic fields must be a few $10^9$~G for WDs with periods around $10^3$~s. For the only LPT with measured spindown rate (not an upper limit), CHIME J0630+25 has an estimated magnetic field of $9.7\times10^9$~G from dipole braking of a WD and is expected to lie above the death line. Moreover, the magnetospheric structure may be complex, exhibiting small-scale multipolar structure. As clarified above (see discussion around Fig. \ref{fig:deathlineWD}), the small $r_c$ values are interpreted as local curvature in a multipolar surface field, not as the curvature of a global dipole. In other words, the WDs would require a complex, intense multipolar field with a small $r_c$. For higher-order multipoles, such as quadrupolar and octupolar fields, the strength of $B_s$ becomes significantly higher than the inferred dipole field, and the geometry of the field becomes much more convoluted. The curvature radius can be even smaller than the stellar radius in a highly localized spot. Previous work has shown that very small radii of curvature, i.e., small-scale sunspot-like fields, can shift the curvature radiation death line to much lower surface magnetic field values \citep{gil/2001}. A modest offset of the dipole, which creates an asymmetric polar cap, can also dramatically enhance pair production \citep{2011ApJ...726L..10H}. Also, LPTs like  GLEAM-X J1627--5235 likely represent a specific ultra-massive subset of WD population, because the death line magnetic field requirement scales with $R^{-21/8}$. Using the radius of a massive WD (2500 km) instead of a standard one (6000 km) reduces the required magnetic field strength by nearly one order of magnitude for the same period. In short, realistic field geometry and acceleration physics can markedly relax the death-line cutoff. We should consider the above as estimates since more reliable values of the death line for WD-pulsars require the nontrivial analysis of the possible effects by the magnetic axis inclination relative to the rotation axis, magnetic axis shift from the center, the polar cap size, magnetic field geometry, the charged particle density flow, general relativity, and additional complications~ \citep[see, e.g.,][, and references therein]{2011ApJ...726L..10H,2016MNRAS.457.2401C,2019ApJ...887L..23B,2019MNRAS.485.4573P,2020ApJ...889..165D,2024JHEAp..42...52D,2021ApJ...920...57C,2022MNRAS.510.2572B,2022MNRAS.516.5084B}, the death-line is not a universal boundary, but is highly sensitive to the stellar radius and curvature of the magnetic field. Independently, it is worth emphasizing that these simple estimates yield reasonable estimates for massive, fast-rotating, highly magnetized WDs, including a light cylinder radius comparable to the star's size. On the other hand, applying the same model to a neutron star leads to ultrastrong magnetic fields of $10^{16}$--$10^{17}$ G and a light cylinder radius a million times the star radius, which seem not to fit within a reasonable picture~\citep[see, e.g., discussion in][]{2023Natur.619..472K}. Furthermore, \cite{2024ApJ...961..214R} notes that for NSs, population synthesis cannot explain the number of ultra-long period sources without unrealistic rates of zero magnetic field decay.

Massive WDs ($>$1.3~M$_{\odot}$) are unlikely to form from single-star evolution due to the mass-loss constraints and are almost certainly the products of double white dwarfs mergers or significant binary interaction. Simulations of binary WD mergers show that the process creates a central core surrounded by a hot, differentially rotating envelope and a massive debris disk. Differential rotation is the primary engine for dynamo application. Magnetic fields generated by dynamos in a turbulent post-merger environment are inherently complex and fragmented, and such dynamos naturally distribute magnetic power across many multipolar modes rather than focusing it solely in dipole mode \citep{castro-tapia/2026}. In the post-merger cooling phase, magnetic fields can undergo Hall drift or re-merger from the crust, potentially creating small-scale, intense ‘spots’ of magnetic field at the polar caps. These spots are physical manifestations of the multipolar components required to lower the curvature radius.
The evidence of the complexity of $B$ is not limited to theory. Multi-wavelength studies of millisecond pulsars and magnetars have frequently found that a pure dipole field fails to explain pulse profiles and X-ray hotspots, requiring multipolar surface components to fit data \citep{lockhart/2019}. The transient nature of  GLEAM-X J1627--5235 (active for months) versus GPM J1839--10 (active for decades) suggests a highly variable magnetospheric state. The small-scale multipolar structures that enable pair production may be subject to Hall-driven evolution or reconnection, which can periodically quench or ignite the radio emission. Due to this cycle, population synthesis, such as demonstrated in \cite{2024ApJ...961..214R}, would find it difficult to explain emission, since it models a steady-state dipole rather than a dynamic, multipole-dominated system.

Our model predicts rotational ages of 390 Myr and 572 Myr for GPM J1839–10 and  GLEAM-X J1627--5235, and the consistency with cooling age constraint for $1.33 M_\odot$ WD larger than $556$ Myr supports multi-wavelength constraints, as follows in the next sections. It demonstrates that the post-merger WD, once it has slowed to the observed rotational period, could have cooled sufficiently to be undetectable in optical observations, indicating that the specific sub-population of the merger satisfies all current multi-wavelength constraints.


\section{Minimum Age of GLEAM-X J1627--5235 from Optical Upper Limits}\label{sec:4}

Optical observations provide crucial insights into the origin and nature of compact objects. For instance, optical spectroscopy can confirm a merger origin, as demonstrated by WD J0551+4135, which exhibits carbon and hydrogen absorption lines~\citep{Hollands/2020}. Zeeman splitting of these lines allows estimation of the surface magnetic field, while polarimetry observations constrain the dipole field strength and geometry~\citep[e.g.,][]{2017NatAs...1E..29B,2022MNRAS.510.2998D}. Unfortunately, such data are unavailable for the two sources analyzed here. However, upper limits on GLEAM-X J1627--5235 can be used to constrain the object's age under a WD model.

\citet{2025MNRAS.538..925L} report a $5\sigma$ upper limit of 24.4~mag from VLT/MUSE observations for GLEAM-X J1627--5235. Although upper limits in other bands exist~\citep[see][]{2022ApJ...940...72R}, the VLT/MUSE data provide the strongest constraint, so we focus on them for the age estimate. As noted by \citet{2025MNRAS.538..925L}, the MUSE wavelength range broadly overlaps with the {\it Gaia} $G$ band, except for a small gap between $4000$--$4700$~\AA. We therefore adopt this upper limit as the $G$-band magnitude ($m_G$), allowing straightforward use of {\it Gaia} passbands in WD cooling models. To derive the minimum age, we employ cooling models for massive WDs: for $M > 1.30$~M$\odot$, we use relativistic carbon-oxygen curves from \citet{2023MNRAS.523.4492A}, while for $1.10 < M < 1.30$~M$\odot$, we adopt the H-rich, $Z=0.001$ curve from \citet{2022MNRAS.511.5198C}, chosen to minimize discontinuities at $1.30$~M$_\odot$.

We estimate the V-band extinction ($A_V$) using two complementary approaches: (1) via empirical relations between dispersion measure (DM), photoelectric absorption column density ($N_H$), and $A_V$; and (2) by extracting $A_V$ directly from a 3D extinction map~\citep[e.g.][]{2023MNRAS.524.1855Z} using the assumed distance (estimated from DM). Both methods rely on the connection between distance, DM, $N_H$, and $A_V$. By tracing two distinct pathways, DM~$\rightarrow$~distance~$\rightarrow$~$A_V$, and DM~$\rightarrow$~$N_H$~$\rightarrow$~$A_V$, we can cross-check our extinction estimates.

For the first approach, we use the relation between DM and $N_H$~\citep{2013ApJ...768...64H}:
\begin{equation}
N_H = (0.30^{+0.13}_{-0.09},\mathrm{DM}) \times 10^{20}~\mathrm{cm}^{-2},
\end{equation}
and the correlation between $N_H$ and $A_V$~\citep{2017MNRAS.471.3494Z}:
\begin{equation}
A_V = ( 4.81 \pm 0.05) \times 10^{-22} N_H.
\end{equation}

For GLEAM-X J1627–5235, with $\mathrm{DM} = 57\pm 1$~pc~cm$^{-3}$~\citep{hurley-walker/2022}, we find $A_V = 0.82^{+0.04}_{-0.03}$. Using the second method, assuming a distance of $1.3\pm 0.5$~kpc \citep{hurley-walker/2022} and the extinction map of \citet{xiangyu2024dust}, we find $A_V \sim 0.9$. Unless the source lies behind a local dust cloud, the agreement between the two methods indicates that the DM-based estimate provides a reasonably constrained value. For simplicity, we adopt the first method (DM $\rightarrow N_H \rightarrow A_V$) for subsequent calculations, as it simplifies uncertainty propagation by making $A_V$ independent of distance.

Finally, we estimate $A_G$, the extinction in the $G$ band, as $A_G = 0.87 A_V$~\citep{1989ApJ...345..245C}. The apparent $G$-band magnitude $m_G$ is then:
\begin{equation}
m_G = M_G + 5\log_{10}(d_{\mathrm{pc}}) - 5 + A_G.
\end{equation}

\begin{figure}[t]
    \centering
    \includegraphics[width=0.90\linewidth]{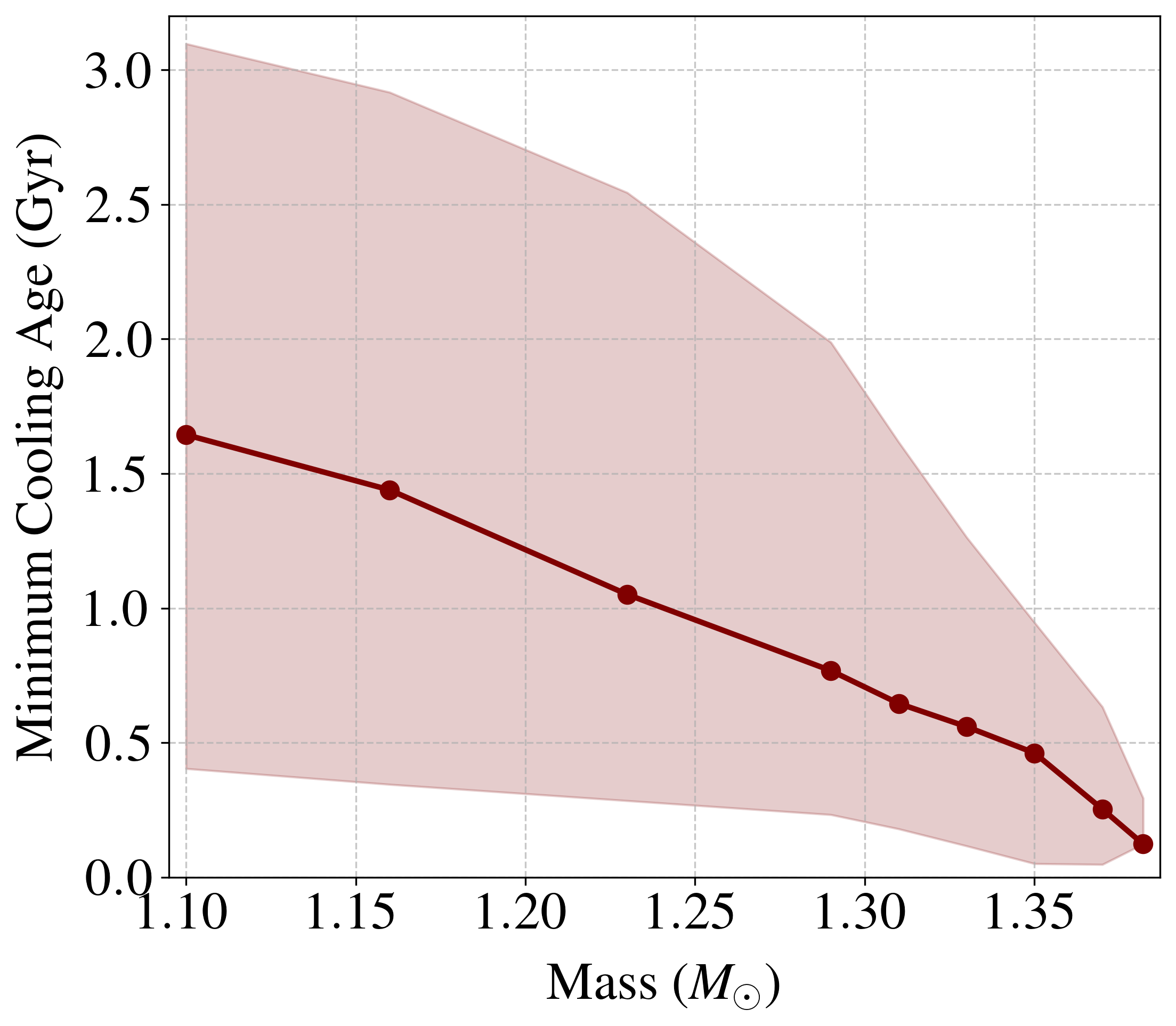}
    \caption{Curve of the minimum age from the optical upper limits for each mass between 1.10 and 1.382 M$_{\odot}$ (blue) for GLEAM-X J1627--5235. The shaded area reflects uncertainties arising from distance and extinction.}
    \label{fig:cooling}
\end{figure}

By comparing $M_G$ from the cooling models with the observed limit $m_G = 24.4$~mag, we estimate the cooling age for masses between 1.10 and 1.382~$M_\odot$. Figure~\ref{fig:cooling} shows that the minimum age for GLEAM-X J1627--5235 ranges from 100~Myr (1.382~$M_\odot$) to 1.6~Gyr (1.10~$M_\odot$), not including uncertainties, which are shown in the figure only. This age is consistent with recently discovered fast massive WDs like J2211+1136~(2.61–2.85 Gyr) and J1901+1458~\citep[10–100 Myr,][]{2022ApJ...941...28S}.

\section{Post-merger rotational evolution}\label{sec:5}

\begin{table*}
\centering
\caption{Values of the initial and derived parameters used in the WD spin-evolution model: mass ($M$), radius ($R$), magnetic field ($B$), accretion rate ($\dot{M}$), disk mass ($M_d$), equilibrium period ($P_{\rm eq}$), time elapsed in \textit{phase I} ($\Delta t_{\rm I}$), \textit{phase II} ($\Delta t_{\rm II}$) and \textit{phase III} ($\Delta t_{\rm III}$), total rotational evolution time ($\Delta t_{\rm evo}$) and spindown rate at the end of evolution ($\dot{P}_{\rm evo}$)}
\begin{tabular}{lrr}
\hline
\multicolumn{1}{c}{\textbf{Parameter}} & \multicolumn{1}{c}{\textbf{\begin{tabular}[c]{@{}c@{}} \\ GPM J1839-10 \\{ }
 \end{tabular}}} & \multicolumn{1}{c}{\textbf{\begin{tabular}[c]{@{}c@{}}GLEAM-X J1627--5235 \end{tabular}}} \\ \hline

\multicolumn{3}{c}{\textbf{\begin{tabular}[c]{@{}c@{}} Initial Parameters \end{tabular}}} \\ \hline
$M$ ($M_\odot$) & 1.33 & 1.33 \\
$R$ (km) & 2500 & 2500 \\
$B$ ($10^{8}$~G) & 10.0 & 10.0 \\
$\dot{M}$ ($10^{-7} M_{\odot}$ yr$^{-1}$) & 2.00 & 3.35 \\
$M_d$ ($M_{\odot}$) & 0.30 & 0.30 \\ \hline

\multicolumn{3}{c}{\textbf{\begin{tabular}[c]{@{}c@{}} Derived Parameters \end{tabular}}} \\ \hline
$P_{\rm eq}$ (s) & 1284.8 & 1030.0 \\
$\Delta t_{\rm I}$ (kyr) & 4.50 & 3.51 \\
$\Delta t_{\rm II}$ (Myr) & 1.495 & 0.892 \\
$\Delta t_{\rm III}$ (Gyr) & 0.388 & 0.571 \\
$\Delta t_{\rm evo}$ (Gyr) & 0.390 & 0.572 \\
$\dot{P}_{\rm evo}$ ($10^{-15} \sin^2\theta$ s s$^{-1}$) & 2.73 & 3.30 \\ \hline
\end{tabular}
    \label{tab:parevo}
\end{table*}

\begin{figure*}
\centering
\includegraphics[width=13.0cm, height=7.0cm]{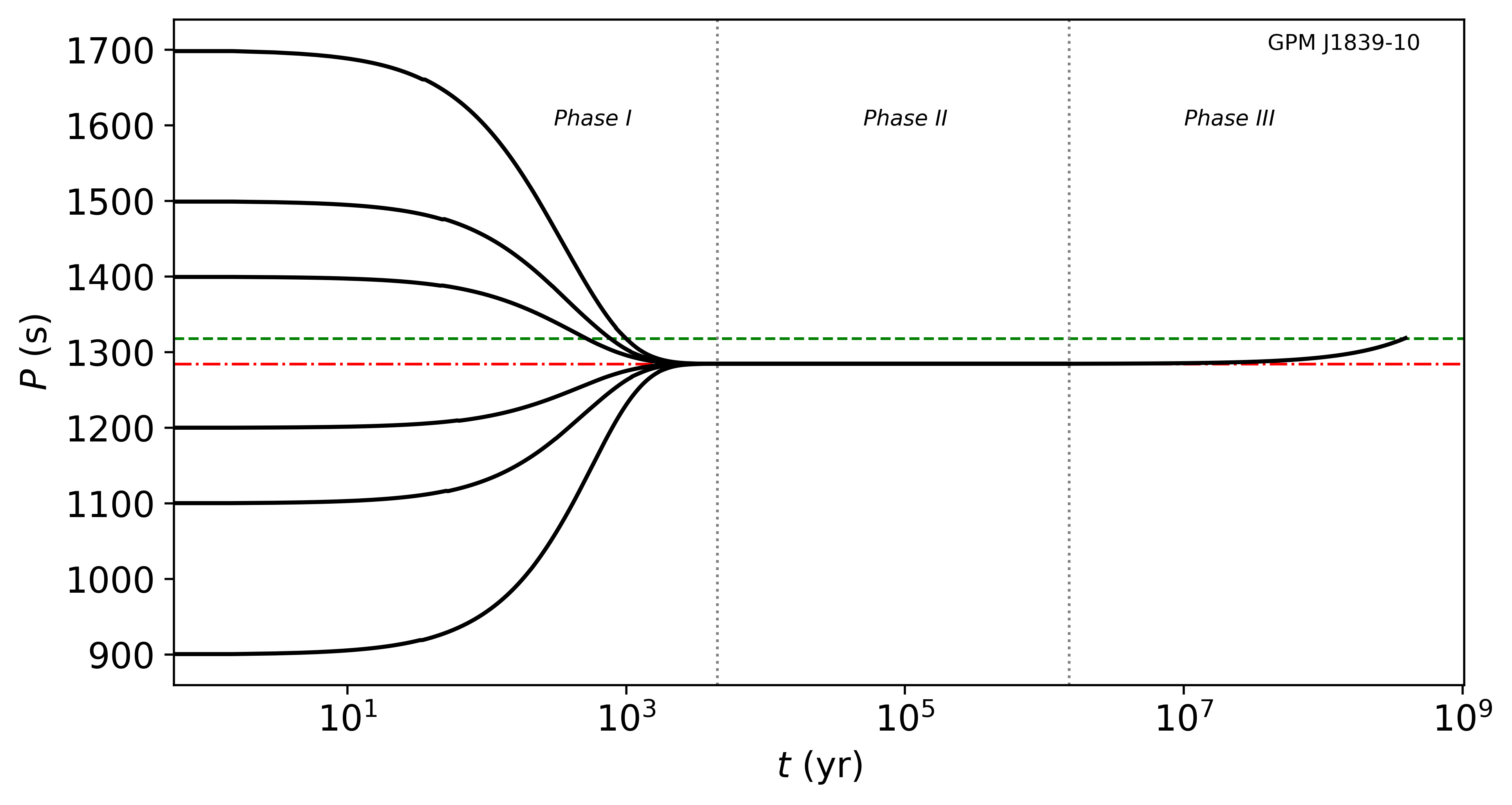}
\caption{Evolution of the rotation period of GPM J1839--10 for an accretion rate of $\dot{M} = 2.0 \times 10^{-7} M_{\odot}$~yr$^{-1}$ and for different values of initial rotational periods, $P_{0} = (900, \, 1100, \, 1200, \, 1400, \, 1500, \, 1700)$~s. The dotted lines divide the evolution into three stages according to the value of $\omega$. The red dash-dotted line indicates the equilibrium period  $P_{\rm eq} = 1284.8$~s. The green dashed line indicates the current rotation period of the WD. } \label{fig:EvoGPM}
\end{figure*}

The merger of two WDs can lead to various outcomes, mainly determined by the masses and compositions of the binary components. These include SN~Ia, NSs formed by collapse, and several classes of hydrogen-deficient stars. A schematic overview of the possible outcomes as a function of WD masses is presented in \cite{2014MNRAS.438...14D}. Among these channels, mergers of two CO WDs are expected to be the most likely pathway to form stable, massive, rapidly rotating, and strongly magnetic WDs in the merger aftermath. Mergers involving oxygen-neon WDs are unlikely to produce a stable massive WD, as their pre-merger masses already exceed $\sim1\,M_\odot$ and only a small amount of mass is expected to be ejected during the event \citep{2009A&A...500.1193L}. Therefore, this scenario most likely forms an NS, given that a long-lasting stable WD remnant would require a low-mass secondary. Mergers between a CO WD and a helium (He) WD may instead produce extreme helium stars or R~Coronae~Borealis stars \citep{2014MNRAS.445..660Z}, while double He WD mergers are thought to form hot subdwarfs \citep{2009ARA&A..47..211H}.

However, not always will CO+CO WD mergers form stable massive WDs, since the outcome depends on the masses of the original WDs. Mergers in which the total mass exceeds the Chandrasekhar limit are generally expected to evolve toward either an SN Ia~\citep{2018ApJ...857..134B,2022ApJ...925...92N} or NS~\citep{2016MNRAS.463.3461S,2018ApJ...857..134B}. Previous works show that for systems with sub-Chandrasekhar total mass, the merger of a mass-symmetric binary can end as an SN~Ia~\citep{2010ApJ...722L.157V}, while, in the case of mass-asymmetry, the secondary is tidally disrupted. In the latter case, the final outcome will be a central WD remnant with a mass approximately equal to the mass of the primary that remains undisturbed, surrounded by a hot envelope made up of approximately half of the secondary. This envelope is surrounded by a rapidly rotating Keplerian disk that contains most of the remaining mass of the less massive WD~\citep[see, e.g.,][]{2009A&A...500.1193L}. The envelope is rapidly accreted onto the central WD, resulting in a uniformly rotating WD surrounded by a Keplerian disk. In this work, we explore the evolution of this latter system.

We now investigate GPM J1839--10 and GLEAM-X J1627--5235 as merger products. Those two LPTs were chosen as examples, but this model can be applied to other LPTs from Table~\ref{tab:LPT_WDs}. For this task, we calculate their post-merger rotational evolution using the recent model presented by \cite{2022ApJ...941...28S}. We examine the evolution of rotational dynamics of the central remnant WD assuming accretion and propeller torque, $T_{\rm acc}$, and magnetic torque, $T_{\rm mag}$ \citep{1999ApJ...520..276M, 2015MNRAS.450..714P}:

\begin{equation}\label{eq:TaccTmag}
    T_{\rm acc} = \dot{M} \sqrt{G M R_m} \left(1-\omega \right),
    \quad T_{\rm mag} = -\frac{2}{3} \frac{B^{2} R^{6} \Omega^{3}}{c^{3}} \sin^{2}\theta,
\end{equation}

\noindent where $\theta$ is the inclination angle of the magnetic dipole moment ($\mu = B R^3$) relative to the WD rotation axis, $R_m$ is the magnetosphere radius given by the Alfvén radius \citep[see, e.g.,][]{1972A&A....21....1P}
\begin{equation}\label{eq:Rm}
    R_m = \left( \frac{\mu^2}{\dot{M} \sqrt{2 G M}}  \right)^{2/7},
\end{equation}
\noindent and $\omega = \Omega/\Omega_K$ is the \textit{fastness} parameter being $\Omega_K = \sqrt{(G M)/(R_m^3)}$ the Keplerian angular velocity at the radius $R_m$. When $\omega < 1$, the material accretes on the stellar surface and transfers angular momentum to it. When $\omega > 1$, the system enters the so-called propeller regime, in which the WD loses angular momentum because its centrifugal barrier ejects the infalling mass from the disk. We consider a constant value of $M$ given by the observed WD mass, and assume it to be the mass of the central, uniformly rotating WD remnant. As mentioned above, to this mass contribute the mass of the primary, of the hot envelope, and a fraction of fallback material. As we shall see, the accreted mass from the disk during the post-merger evolution, $M_{\rm acc}$, is relatively small compared to the total WD mass, so it can be neglected for our purpose. We also assume a constant mass accretion rate, $\dot{M}$, onto the central WD. These simplifications do not significantly affect our results, since both mass accretion and mass ejection deplete the disk on timescales much shorter than the WD lifetime.


The WD rotation evolves differently depending on the model parameters and initial conditions. Generally, it evolves through three stages before reaching its current rotation period, depending on the fastness parameter value: $\omega > 1$, $\omega \approx 1$, or $\omega < 1$. The initial angular velocity, $\Omega_0$, sets the initial fastness parameter in \textit{phase I}: $\omega_0 >1$ ($\Omega_0 > \Omega_K$; propeller removes angular momentum) or $\omega_0 <1$ ($\Omega_0 < \Omega_K$; accretion transfers angular momentum). When $\omega \approx 1$ (\textit{phase II}), the accretion and magnetic torques balance, $T_{\rm acc} \approx T_{\rm mag}$, so the angular velocity remains constant at the equilibrium value, $\Omega \approx \Omega_{\rm eq} = \Omega_K$. After the disk is exhausted, the WD evolves in the $\omega < 1$ regime (\textit{phase III}), with only the magnetic dipole exerting torque.

\begin{figure*}
\centering
\includegraphics[width=13.0cm, height=7.0cm]{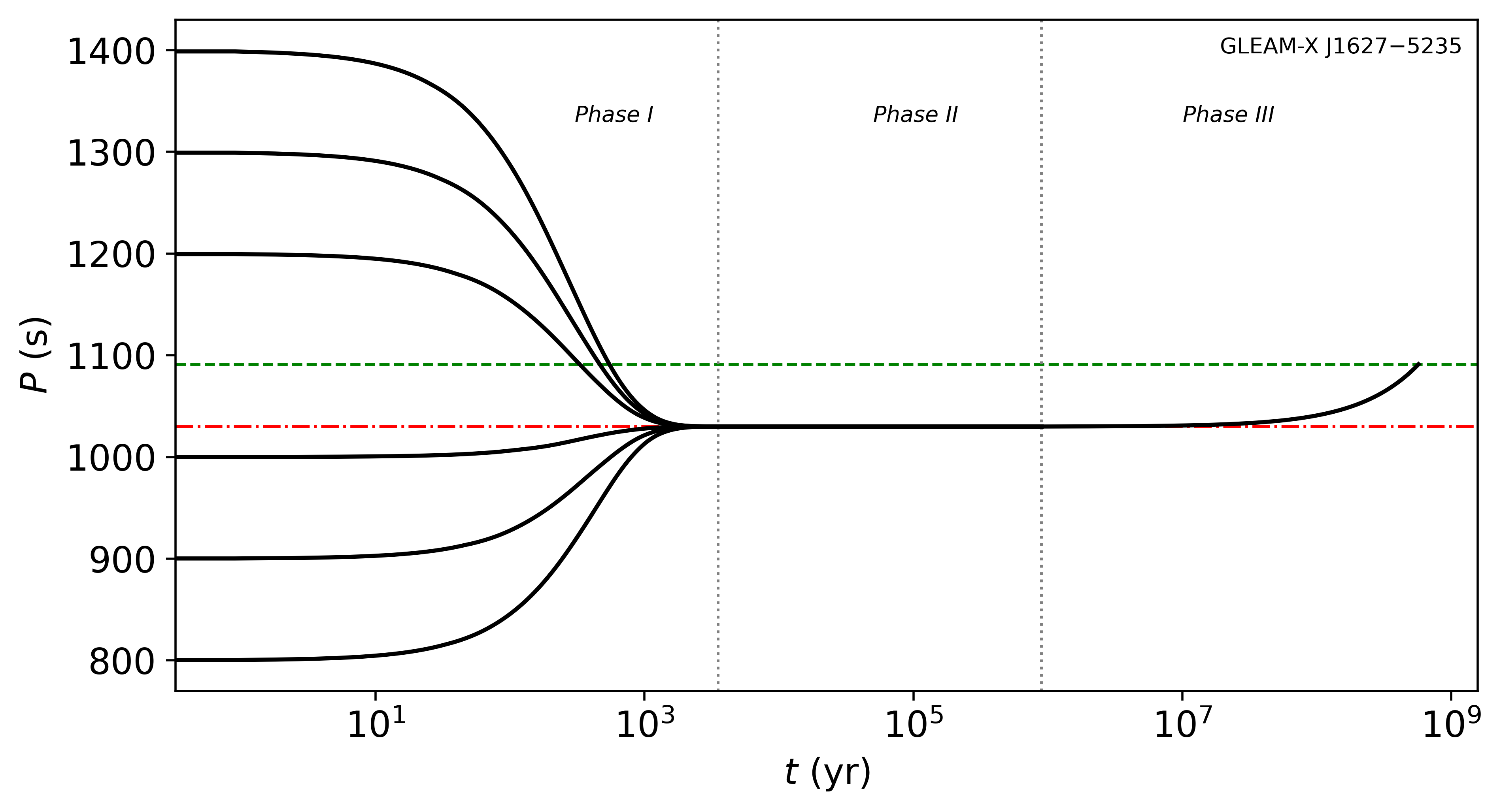}
\caption{Evolution of the rotation period of GLEAM-X J1627--5235 for an accretion rate of $\dot{M} = 3.35 \times 10^{-7} M_{\odot}$~yr$^{-1}$ and for different values of initial rotational periods, $P_{0} = (800, \, 900, \, 1000, \, 1200, \, 1300, \, 1400)$~s. The dotted lines divide the evolution into three stages according to the value of $\omega$. The red dash-dotted line indicates the equilibrium period  $P_{\rm eq} = 1030.0$~s. The green dashed line indicates the current rotation period of the WD.} \label{fig:EvoGlea}
\end{figure*}

Figures \ref{fig:EvoGPM} and \ref{fig:EvoGlea} show the rotation evolution for the two sources, GLEAM-X J1627--5235 and GPM J1839--10, taking into account the initial parameters presented in Table~\ref{tab:parevo} and for $\theta = 90^\circ$. Each phase of evolution described above is identified in this figure. Furthermore, we considered $6$ values for the initial rotation period, $3$ values below and $3$ above the equilibrium period $P_{\rm eq} = 2\pi/\Omega_{\rm eq}$. We note that the solution is not sensitive to this initial condition.

The accretion rate adopted in our calculations is $\dot{M} \sim 10^{-7}~M_{\odot}$~yr$^{-1}$. This value is not arbitrary: it ensures that the rotational evolution remains consistent with the sources estimated ages. It corresponds to approximately $10\%$ of the Eddington accretion rate and lies within the range expected for double WD mergers. Hydrodynamical simulations of such systems \citep[e.g.,][]{2015ApJ...805L...6S, 2016MNRAS.463.3461S, 2014MNRAS.438...14D} show that, once the secondary is disrupted, the accretion of the debris disk onto the primary proceeds at near- or super-Eddington rates. Specifically, immediately after the dynamical merger phase, the mass transfer rate can reach $\dot{M} \gtrsim 10^{-5}~M_{\odot}$~yr$^{-1}$ \citep[see also][]{2007MNRAS.380..933Y, 2018ApJ...857..134B}. During the subsequent viscous evolution, however, the accretion rate decreases and stabilizes in the range $10^{-5}$–$10^{-7}\, M_{\odot},\mathrm{yr}^{-1}$ \citep{2012MNRAS.419..452Z}, consistent with the value required in our model.

Table~\ref{tab:parevo} also presents the parameters derived from the evolution simulation. We observe that the elapsed time required for GPM~J1839--10 and GLEAM-X~J1627 to reach the current measured period is $\Delta t_{\rm evo} \approx 0.390$~Gyr and $\Delta t_{\rm evo} \approx 0.572$~Gyr, respectively. The age for GLEAM-X~J1627 agrees with the optical upper limits that predict an age $>0.556$~ Gyr for a $1.33\,M_\odot$ WD.  Another interesting parameter worth mentioning is the spindown rate $\dot{P}_{\rm evo}$ at the end of the spin evolution, that is, the $\dot{P}_{\rm evo}$ value that must be observed in the current phase of the source given that only the magnetic torque acts on the star. For the values in Table~\ref{tab:parevo}, we infer spindown rates $\sim 10^{-15}$ s s$^{-1}$ for both sources.

\section{Limiting Magnitudes for WD models} \label{sec:6}

An important way to confirm or rule out the presence of a WD is to look for an optical counterpart that matches the WD's photospheric emission. \citet{2025MNRAS.538..925L} note that the current deepest upper limit in the {\it Gaia} $G$ band ($\sim$24.4~mag) for GLEAM-X J1627–5235 lies close to the expected brightness range of WDs at this distance, typically $m_G \sim 22$–24~mag. However, as shown in Sec.~\ref{sec:4}, this limit does not strongly constrain massive WDs, even for ages below 1~Gyr.

\begin{figure}
    \centering
    \includegraphics[width=0.95\linewidth]{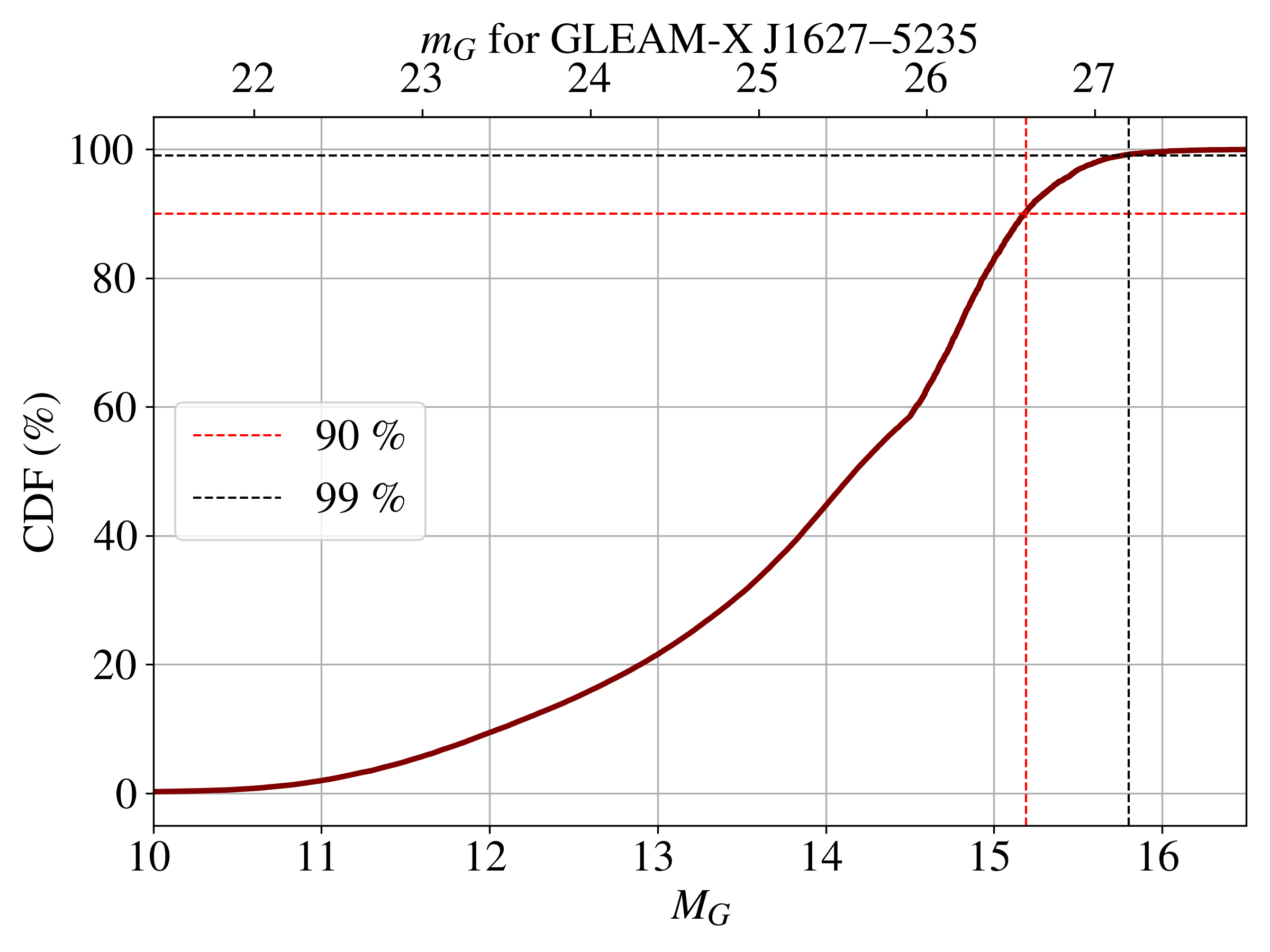}
    \caption{CDF of {\it Gaia} $M_G$ absolute magnitudes for WDs (bottom x-axis). The top x-axis shows the corresponding apparent $G$-band magnitudes for GLEAM-X J1627–5235, considering its estimated distance and extinction. The red and black dashed lines mark the magnitudes below which 90\% and 99\% of the WD population lie, respectively.}
    \label{fig:cdf}
\end{figure}

To quantify this, Figure~\ref{fig:cdf} shows the cumulative distribution function (CDF) of {\it Gaia} WDs (all masses) within 200 pc \citep{2021MNRAS.508.3877G}, with contributions from fainter WDs corrected for incompleteness following the procedure of \citet{2022MNRAS.511.5984F}. Specifically, WDs fainter than {\it Gaia's} survey limit are not detectable throughout the full 200 pc volume; therefore, each star is weighted by $V_{\rm max}/V_{\rm lim}(M_G)$, where $V_{\rm lim}(M_G)$ is the maximum volume within which WDs of absolute magnitude $M_G$ remain detectable.

This analysis shows that the current limit of $m_G \sim 24.4$~mag corresponds roughly to the 25th percentile of the WD brightness distribution. Reaching the 90th percentile would require sensitivity to $m_G \sim 26.5$~mag, well beyond the current upper limits. Furthermore, GLEAM-X J1627--5235 may host an unusually massive WD. For example, a $1.382$~M$_\odot$ WD with an age of 750~Myr would have $m_G \sim 28.5$~mag, even fainter than the percentile thresholds. Thus, confirming an WD origin for GLEAM-X J1627--5235 may require optical sensitivities achievable with next-generation telescopes (e.g., ELT and GMT). We note that the estimated age in Sec.~\ref{sec:5} depends on the assumed $\dot M$ and thus could be adjusted to accommodate a cooler, and hence older, WD if future observations identify one.

For GPM J1839--10, a candidate $z_s$-band counterpart of 22.7~mag has recently been reported \citep{2025MNRAS.544L..76P}, although its association remains unconfirmed. If the counterpart is real, whether the system is an isolated WD or a WD+M-dwarf binary (e.g., ILT J1101+5521), the inferred distance of $\sim5.7$ kpc is almost certainly overestimated. At such a distance, an optical detection would be highly unlikely with current facilities. This source has ${\rm DM}=273.5$ pc cm$^{-3}$, corresponding to $A_G\sim 3.4$~\citep{hurley-walker/2023}. Even if the source were an intrinsically bright WD with absolute magnitude $M_G \sim 10$, its apparent magnitude would be $m_G \sim 27$. 

For ASKAP~J1935+2148, the expected apparent magnitude of a bright WD ($M_G\sim 10$) is $m_G\sim 25$, while for a WD at the $95^{\rm th}$ percentile of the luminosity distribution ($M_G\sim 15.5$) the expected magnitude is $m_G\sim 30.5$. If this object is a young, bright WD, it may be observable with current telescopes; however, ruling out an isolated WD scenario will require future facilities.

For CHIME~J0630+25, its relatively small distance and dispersion measure imply $m_G\sim 22$ even at the $95\%$ confidence level, and thus it should be observable. In this case, as noted by \cite{2025ApJ...990L..49D}, establishing a confident association is difficult because the field is crowded and the radio localization is poor, leaving roughly $\sim150$ marginal candidates consistent with the position and period. More precise localization is required to confidently identify a counterpart. For DART~J1832$-$0911 and ASKAP~J1755$-$2527, even a bright WD would have expected magnitudes of $m_G\sim 29$ and $m_G\sim 32$, respectively, making it unlikely that their nature can be determined with current optical facilities.

\section{Discussion and Conclusion }\label{sec:7}

In a polar cap model context, we showed that the request that isolated LPTs lie above the death line for WD-pulsars requires magnetic fields of a few $\sim 10^9$ G and the presence of small-scale multipolar structure in the magnetosphere (see Section \ref{sec:3}). The small-scale multipolar magnetosphere is not merely an auxiliary assumption but a necessary consequence of the merger origin. Differential rotation in the post-merger debris is a robust mechanism for generating non-dipolar surface fields, which are empirically observed in other classes of compact objects. We also include in the analysis a set of observed isolated WDs that are confirmed to be massive, fast-rotating, and highly magnetized. The analysis also suggests the mass ($\sim 1.3\,\text{M}_\odot$) and radius ($\sim 2500$ km) for the LPTs in Table~\ref{tab:LPT_WDs}, indicating that they are consistent with being made of carbon or oxygen, the most likely compositions for post-merger massive WDs. 

We also touched on several pieces of observational evidence of massive and highly magnetized WDs, including the prevalence of WD binaries compared to solar-type main-sequence star binaries. This suggests that a significant fraction of isolated WDs may be double WD merger products. This idea has long been suggested, and the remnants of these mergers can be stable massive WDs, SN Ia, hydrogen-deficient stars (such as R Coronae Borealis stars), or NSs. Recent observations support this hypothesis, with rapidly rotating, strongly magnetized, and massive WDs potentially arising from mergers of WD binaries. Therefore, we have examined the hypothesis that massive WDs can result from merging WD binaries, thereby providing a merger origin for some LPTs. More specifically, it can be an alternative to the WD+M-Dwarf interpretation, specially for objects such as GLEAM-X J1627--5235 that lacks an optical counterpart consistent with the presence of such binary systems. We performed the post-merger spin evolution using a model presented in a previous study~\citep{2022ApJ...941...28S} for two LPTs, GLEAM-X J1627--5235 and GPM J1839--10, which explains how the stars' angular momentum changes due to accretion, propeller, and magnetic torques. From the modeling, we inferred the rotational ages of both sources, i.e., the time required for them to reach their currently observed rotational periods. We obtained $572$ Myr and $390$ Myr for GLEAM-X J1627--5235 and GPM J1839--10, and verified that the upper limit on their spindown rate is also satisfied (see Section \ref{sec:4} for details).

We also note that this model is consistent with potential X-ray observations, given that accretion episodes from the disk are expected to happen. \cite{Borges2020} noted that a WD accreting matter from the disk can have emission in the soft and hard X-rays, coming from the post-shock region and the hot spots in the WD surface. One of the LPTs present in Table~\ref{tab:LPT_WDs}, DART~J1832-0911, has detected X-ray emission, with a luminosity of $\sim 10^{33}$~erg.s$^{-1}$~\citep{2025Natur.642..583W}. In this case, we can check the expected accretion rate for such luminosity. A good order of magnitude of the accretion rate is $GM \dot M/R$, and considering the mass and radius of table~\ref{tab:LPT_WDs}, we find $\sim10^{-11}$~M$_{\odot}$.yr$^{-1}$. This value is consistent with that of other accreting WD classes, such as intermediate polars. Thus, a post-merger WD is consistent with the current paradigm that some, but not all, LPTs exhibit episodic X-ray emission, given that the emission may be linked to the amount of gas remaining in the disk.

Theoretical and observational work on WD pulsars and related areas of physics and astrophysics is flourishing and poised for breakthroughs, given the advanced capabilities of new observational facilities and theoretical advances. Their pulsar-like emission might be observed as magnetar-like activity
~\citep[see, e.g.,][ and references therein]{2013ApJ...772L..24R,2017A&A...599A..87C,2018ApJ...857..134B,Borges2020}. An infrared/optical/ultraviolet transient from the cooling of the expanding merger dynamical ejecta peaks a few hours post-merger~\citep{2022IJMPD..3130013R,2019JCAP...03..044R,2023ApJ...958..134S}, and is ready for observation by the forthcoming Vera Rubin Observatory at a predicted detection rate of up to 1000 a year \citep{2023ApJ...958..134S}.  Additional multimessenger theoretical predictions involve the emission and possible detection of high-energy neutrinos \citep{2016ApJ...832...20X}, cosmic rays~\citep{2011PhRvD..83b3002K,2017JPhCS.861a2005L}, gamma-rays~\citep{1988SvAL...14..258U,2023Galax..11...14M,2006A&A...445..305I}, and the gravitational waves by merging double WDs, detectable by \textit{LISA}, the Laser Interferometer Space Antenna,  \citep{Stroeer2006Sep,Korol2022Apr}. \textit{LISA} will possibly measure deviations from pure gravitational-wave-driven orbital dynamics, e.g., given the presence of electromagnetic radiation \citep{2022ApJ...940...90C}. Gravitational waves from quadrupole deformations of WD pulsars might also be detected \citep{2020MNRAS.492.5949S, 2024MNRAS.531.1496S}.
The Rubin/LSST observing cadence (typically revisiting a given field on timescales of a few days) raises the concern that a fast-evolving, kilonova-like optical transient might be missed or poorly sampled. For the double-WD-merger ``kilonova-like'' emission considered in \cite{2023ApJ...958..134S}, the optical/IR light curves could remain detectable over multi-day timescales (with the brightest phase lasting of order a few days, followed by a multi-day decline; see the light curves), even with a 2--3 day cadence, for events above the LSST threshold magnitude. Quantitative light-curve examples and estimated upper limits on the LSST detection rate of the expected optical transient from WD-WD mergers have been presented in \cite{2023ApJ...958..134S}. A dedicated discussion of the effects of cadence and the consequent reduction (relative to the upper limits by \citealp{2023ApJ...958..134S}) in LSST detection rate of the WD-merger optical transient has recently been presented by M. M. Ridha Fathima et al. (private communication; paper submitted).

In summary, we have provided insights into the properties and origin of LPTs, massive WDs with short rotation periods and strong magnetic fields. Our analysis suggests that the carbon or oxygen abundance is most likely for such WDs, with minor differences in maximum mass and radius. The observations support the hypothesis that these massive WDs can result from WD binary mergers. Recent data and the characteristics of these merged remnants, such as strong magnetism and rapid rotation, further substantiate this. We have highlighted the importance of future optical observations of LPT fields to further constrain and confirm the nature of these objects.

The proposition that LPTs are highly magnetized, fast-rotating WDs formed through double WD mergers broadens our understanding of compact objects. Confirming the isolated WD nature for some LPTs would redefine our understanding of double WD merger outcomes, emphasizing their role not only in Type Ia supernovae but also in producing remnants such as WD pulsars. Future observations using upcoming optical facilities (e.g., ELT and GMT) will be crucial for validating these predictions. Upcoming deep optical searches and future multimessenger observations (e.g., LISA) will be decisive in confirming whether LPTs constitute a population of
WD pulsars.


\section*{Acknowledgments}
This work was initiated by Manuel Malheiro and is dedicated to his memory. Manuel was a renowned Brazilian physicist who passed away on June 09, 2024, in S\~ao Paulo, Brazil, during the development of this work. He left behind an indelible legacy for the Brazilian nuclear physics and relativistic astrophysics communities that extends beyond scientific knowledge.\\[1em]

The authors thank the referee for the insightful comments. S.V.B. acknowledges support from the NASA ATP program (NNH23ZDA001N-ATP)
and the College of Letters and Science at UWM through the Chancellor's
Graduate Student Award and the Research Excellence Award. J.G.C. is
grateful for the support of FAPES (1020/2022, 1081/2022, 976/2022,
332/2023), CNPq (311758/2021-5, 306018/2025-0), and FAPESP (grant
No. 2021/01089-1). RVL was supported by INCT-FNA (Instituto Nacional de
Ci\^encia e Tecnologia, F\'isica Nuclear e Aplica\c c\~oes), research
Project No. 464898/2014-5, also thanks CNPq/CAPES, and is very grateful to Professor Manuel Malheiro for his scientific guidance, support, and
partnership over several years. M.F.S. thanks FAPESP (2025/05794-2,
2021/01089-1) for the financial support and for the support of CNPq
(173535/2023-2). E. Otoniel acknowledges support from
FUNCAP(BP6-0241-00335.01.00/25). F.W. is supported through the
U.S. National Science Foundation under Grant PHY-2012152.

This work has made use of data from the European Space Agency (ESA) mission {\it Gaia} \citep[\url{https://www.cosmos.esa.int/gaia},][]{2016A&A...595A...1G, 2023A&A...674A...1G}, processed by the {\it Gaia} Data Processing and Analysis Consortium \citep[DPAC,][ \url{https://www.cosmos.esa.int/web/gaia/dpac/consortium}]{2023A&A...674A..32B}. Funding for the DPAC has been provided by national institutions, in particular, the institutions participating in the {\it Gaia} Multilateral Agreement.

\section*{Data Availability}
The codes used in this study are available from the corresponding author upon reasonable request. All datasets used in this study are publicly available and can be accessed from their original sources:
\begin{itemize}
    \item WD cooling models: \url{https://evolgroup.fcaglp.unlp.edu.ar/modelos.html}
    \item The 3D interstellar extinction map: \url{https://zenodo.org/records/11394477}
    \item The \textit{Gaia} white dwarf catalog: \url{https://warwick.ac.uk/fac/sci/physics/research/astro/research/catalogues/gaiaedr3_wd_main.fits.gz}
\end{itemize}

\appendix

\bibliographystyle{elsarticle-harv}
\bibliography{bibliography}






\end{document}